\documentclass[aps,prmaterials,superscriptaddress,preprint]{revtex4-2}
\usepackage{booktabs}
\usepackage{graphicx}
\usepackage{multirow}
\usepackage{amsmath}
\usepackage{xcolor}
\usepackage{textgreek}
\usepackage{booktabs}
\usepackage{tabularx}
\DeclareUnicodeCharacter{2212}{-}

\draft 

\begin{document}


\title{A Review of AI-Driven Approaches for Nanoscale Heat Conduction and Radiation} 



\author{Ziqi Guo}%
\affiliation{ 
School of Mechanical Engineering, Purdue University, West Lafayette, 47907, IN, USA
}%
\affiliation{ 
The Birck Nanotechnology Center, Purdue University, West Lafayette, 47907, IN, USA
}%

\author{Daniel Carne}%
\affiliation{ 
School of Mechanical Engineering, Purdue University, West Lafayette, 47907, IN, USA
}%
\affiliation{ 
The Birck Nanotechnology Center, Purdue University, West Lafayette, 47907, IN, USA
}%

\author{Krutarth Khot}%
\affiliation{ 
School of Mechanical Engineering, Purdue University, West Lafayette, 47907, IN, USA
}%
\affiliation{ 
The Birck Nanotechnology Center, Purdue University, West Lafayette, 47907, IN, USA
}%

\author{Dudong Feng}%
\affiliation{ 
School of Mechanical Engineering, Purdue University, West Lafayette, 47907, IN, USA
}%
\affiliation{ 
The Birck Nanotechnology Center, Purdue University, West Lafayette, 47907, IN, USA
}%

\author{Guang Lin}%
 \email{guanglin@purdue.edu}
\affiliation{ 
School of Mechanical Engineering, Purdue University, West Lafayette, 47907, IN, USA
}%

\author{Xiulin Ruan}%
 \email{ruan@purdue.edu}
\affiliation{ 
School of Mechanical Engineering, Purdue University, West Lafayette, 47907, IN, USA
}%
\affiliation{ 
The Birck Nanotechnology Center, Purdue University, West Lafayette, 47907, IN, USA
}%

\date{\today}

\begin{abstract}

Heat conduction and radiation are two of the three fundamental modes of heat transfer, playing a critical role in a wide range of scientific and engineering applications ranging from energy systems to materials science. However, traditional physics-based simulation methods for modeling these processes often suffer from prohibitive computational costs. 
In recent years, the rapid advancements in Artificial Intelligence (AI) and machine learning (ML) have demonstrated remarkable potential in the modeling of nanoscale heat conduction and radiation. 
This review presents a comprehensive overview of recent AI-driven developments in modeling heat conduction and radiation at the nanoscale.
We first discuss the ML techniques for predicting phonon properties, including phonon dispersion and scattering rates, which are foundational for determining material thermal properties. 
Next, we explore the role of machine-learning interatomic potentials (MLIPs) in molecular dynamics simulations and their applications to bulk materials, low-dimensional systems, and interfacial transport. 
We then review the ML approaches for solving radiative heat transfer problems, focusing on data-driven solutions to Maxwell’s equations and the radiative transfer equation. 
We further discuss the ML-accelerated inverse design of radiative energy devices, including optimization-based and generative model-based methods.
Finally, we discuss open challenges and future directions, including data availability, model generalization, uncertainty quantification, and interpretability. Through this survey, we aim to provide a foundational understanding of how AI techniques are reshaping thermal science and guiding future research in nanoscale heat transfer.

\end{abstract}

\maketitle 

\section{Introduction}


Heat conduction and radiation are two of the three fundamental modes of heat transfer. In insulators and semiconductors, atomic vibration is the dominant mode of conduction. It is crucial in various of applications including thermal switches~\cite{wehmeyer2017thermal}, building energy savings~\cite{lindsay2018survey,lirecent}, thermal management of semiconductor devices~\cite{moore2014emerging,he2022experimental}, thermal energy storage systems~\cite{agyenim2010review}, thermoelectrics~\cite{zebarjadi2012perspectives}, and thermal barrier coatings~\cite{flamant2019opportunities}. Radiation, present in all matter above absolute zero Kelvin temperatures, involves the transfer of thermal energy via electromagnetic waves. It is important in applications including photovoltaic energy generators~\cite{sampaio2017photovoltaic}, polaritonics~\cite{caldwell2015low}, thermal-photonic devices~\cite{feng2021near,zhang2024modulation}, radiative energy converters~\cite{hu2018feature,feng2022thermoradiative} and radiative cooling~\cite{tong2022electronic}. 
To model these transport phenomena, a variety of physics-based computational approaches have been developed, including density functional theory (DFT) calculations, molecular dynamics (MD) simulations, the Boltzmann transport equation (BTE)~\cite{ziman2001electrons}, and the radiative transfer equation which can be derived from BTE~\cite{modest2021radiative}. While these methods provide a rigorous foundation for understanding energy transport, they are often computationally expensive, especially when applied to large-scale or high-throughput studies.

Artificial intelligence (AI), particularly machine learning (ML), has witnessed remarkable growth in recent years. In the field of computer vision, machine learning algorithms have revolutionized image recognition~\cite{krizhevsky2012imagenet}, enabling computers to categorize visual information with unprecedented accuracy. Similarly, in natural language processing, machine learning techniques have empowered machines to understand and interpret human language~\cite{otter2020survey}, propelling advancements in areas such as chatbots~\cite{adamopoulou2020overview,liu2024deepseek}, translation services~\cite{hirschberg2015advances}, and text mining~\cite{tan1999text,kumar2021applications,li2021text}. Some specific hardware has also been developed to accelerate the training and inferences of ML models~\cite{qin2024high,jouppi2017datacenter,shawahna2018fpga}. The impact of machine learning extends far beyond these domains. ML has emerged as a promising tool to augment or replace traditional physics-based solvers. 
With the ability to learn complex patterns from large datasets and make fast predictions, ML has the potential to overcome the limitations of traditional computational methods.
In nanoscale heat transfer, the motivation for using ML is twofold.
Firstly, ML models can be trained as surrogates for physics simulations, providing fast predictions~\cite{jaafreh2021lattice,li2024integrated}. This could enable tasks such as high-throughput prediction of material properties~\cite{pyzer2022accelerating} and real-time prediction for operating systems, which is too slow with first-principles or numerical solvers. Secondly, ML can efficiently search large design spaces for materials and devices with target thermal properties~\cite{axelrod2022learning,molesky2018inverse}, which is extremely challenging using brute-force methods or human intuition alone.

\begin{figure}[h]%
\centering
\includegraphics[width=\textwidth]{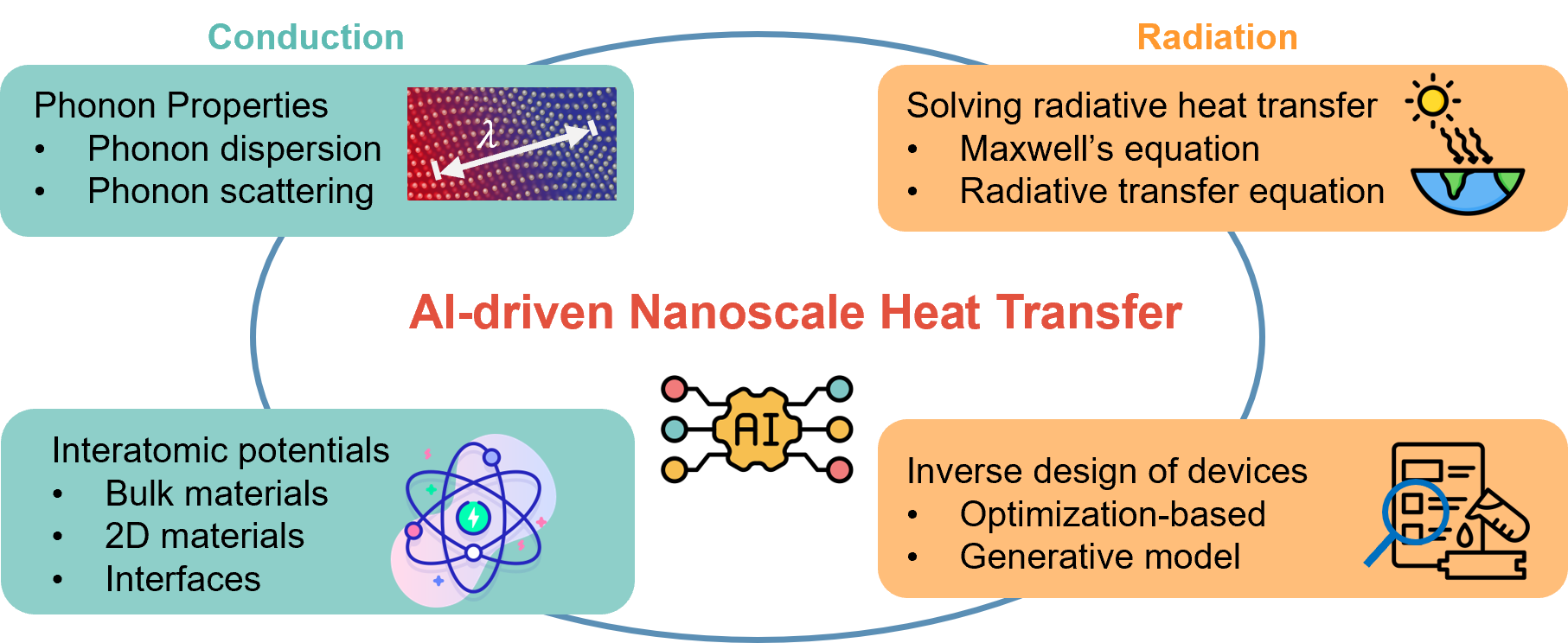} 
\caption{\textbf{Overview of this review.} 
}\label{Fig_overview}
\end{figure}

This review provides a comprehensive overview of AI-driven approaches for nanoscale heat conduction and radiation, as shown in Fig.~\ref{Fig_overview}. \textcolor{black}{The paper is organized into four key sections. It begins by discussing how machine learning is used to predict fundamental phonon properties, such as phonon dispersion and scattering, which are critical for understanding heat conduction in materials. Next, it explores the use of machine learning interatomic potentials to accelerate molecular dynamics simulations, enabling the study of thermal transport in bulk materials, low-dimensional systems, and interfaces with near-first-principles accuracy. The review then shifts to AI approaches for radiative heat transfer, covering data-driven solutions to Maxwell's equations and the radiative transfer equation. Finally, it discusses the use of ML to accelerate the inverse design of thermal radiative devices, including both optimization-based and generative model-based methods.}
In each section, we discuss how machine learning models such as multilayer perceptrons (MLP), graph neural networks (GNN), random forests, diffusion models, and other techniques have been applied in these domains (summarized in Table~\ref{tab:ml_thermal}), and analyze how these models compare to or enhance traditional methods.
This review concludes with a future outlook for AI in nanoscale heat transfer modeling.
Through this survey, we aim to provide a comprehensive overview of the current state of this rapidly growing field.

\begin{table}[h]
\caption{Summary of machine learning models discussed in this review}
\label{tab:ml_thermal}
\begin{tabular}{l|l|l}
\hline
\textbf{Application Area} & \textbf{Problem} &\textbf{ Machine Learning Models/Techniques} \\ \hline
\multirow{2}{*}{Phonon properties} & Phonon dispersion prediction & \begin{tabular}[c]{@{}l@{}}Graph Neural Network\\ Transfer Learning\end{tabular} \\ \cline{2-3}
& Phonon scattering & \begin{tabular}[c]{@{}l@{}}Multilayer Perceptron\\ Random Forest\\ Maximum Likelihood Estimation\end{tabular} \\ \hline
Interatomic Potentials & MD simulations & \begin{tabular}[c]{@{}l@{}}Neural Network Potential \\ Gaussian Approximation Potential\\  Moment Tensor Potential\\  Spectral Neighbor Analysis Method\\ Atomic Cluster Expansion\end{tabular} \\ \hline
Radiative Heat Transfer & Solving Maxwell's equation/RTE & \begin{tabular}[c]{@{}l@{}}Physics-Informed Neural Network\\ Convolutional Neural Network \\ Residual Neural Network\\ Tandem Neural Network\end{tabular} \\ \hline
\multirow{2}{*}{Inverse Design} & Optimization-based approach & \begin{tabular}[c]{@{}l@{}}Bayesian optimization\\ Genetic algorithm\\ Monte Carlo tree search\end{tabular} \\ \cline{2-3}
& Generative model approach & \begin{tabular}[c]{@{}l@{}}Variational Autoencoder\\ Generative Adversarial Network\\ Diffusion Model\\ Tandem Neural Network\end{tabular} \\ \hline
\end{tabular}
\end{table}

\section{ML prediction of phonon properties}

As quantized modes of lattice vibrations, phonons play a central role in heat conduction. Accurate prediction of phonon properties is essential for understanding and designing materials with desired thermal characteristics. State-of-the-art approaches rely on \textit{ab initio} calculations to obtain the harmonic and anharmonic force constants, then solve the phonon BTE for scattering and transport coefficients. These first-principles workflows are accurate but extremely computationally expensive, especially four-phonon scattering. In recent years, a variety of machine-learning approaches have been developed to predict phonon properties more efficiently. 
While several meaningful attempts took an end-to-end approach to predict a material’s lattice thermal conductivity from simple atomic descriptors (including atomic masses, bondings, crystal structure, etc.)~\cite{zhu2021charting,liu2022leveraging,srivastava2023end}, they were limited by data scarcity and have yet to show the accuracy of first-principles level. 
Alternatively, ML may be used to predict at the level of phonon properties, i.e., phonon dispersion curves (frequency vs. wavevector) and phonon lifetimes, which can then be used to compute thermal conductivity. Rather than learning an opaque mapping from structure to conductivity, this approach tries to predict the intermediate phonon properties that feed into transport calculations, which keeps more physics.

\begin{figure}[h]%
\centering
\includegraphics[width=0.6\textwidth]{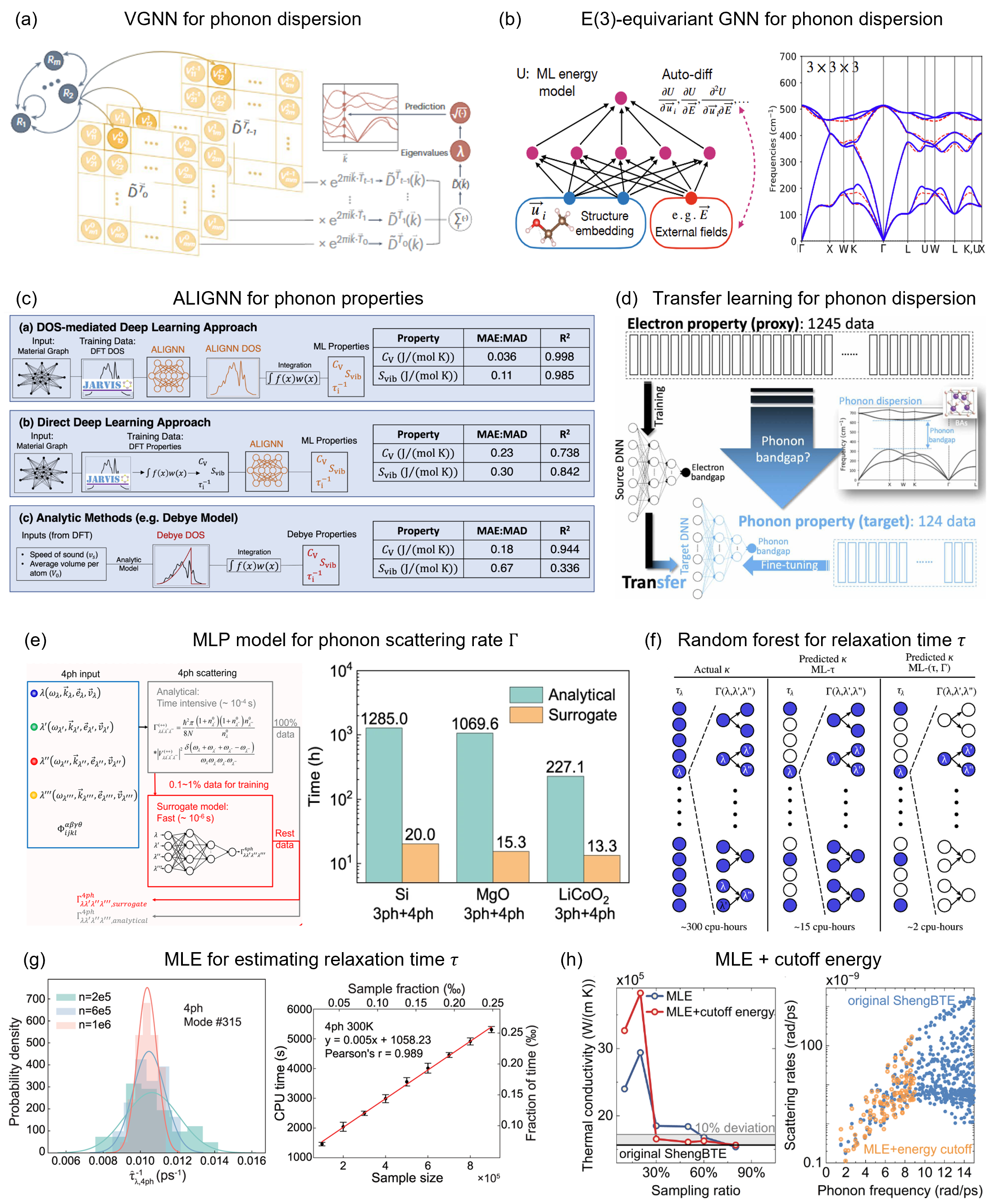} 
\caption{\textbf{ML prediction of phonon properties.} (a) Virtual node GNN for predicting phonon dispersion, as adapted from Okabe et al.~\cite{okabe2023virtualnodegraphneural}, (b) E(3)-equivariant GNN for phonon dispersion prediction, as adapted from Fang et al.~\cite{fang2024phonon}, (c) ALIGNN for predicting phonon properties, as adapted from Gurunathan et al.~\cite{gurunathan2023rapid}, (d) Transfer learning for phonon dispersion, as adapted from Liu et al.\cite{liu2020leverage}, (e) Multilayer perceptron for predicting phonon scattering rate, as adapted from Guo et al.~\cite{guo2023fast}, (f) random forest model for predicting phonon relaxation time, as adapted from Srivastava et al.~\cite{srivastava2024accelerating}, (g) Maximum likelihood estimation (MLE) method for predicting phonon relaxation time, as adapted from Guo et al.~\cite{guo2023samplingacceleratedfirstprinciplespredictionphonon}, (h) Combining MLE method with cutoff phonon frequency, as adapted from Zhang et al.~\cite{zhang2024cryogenic}. 
}\label{Fig_phonon_dynamics}
\end{figure}

\subsection{Phonon dispersion}

Phonon dispersion describes the relationship between phonon frequency and wavevector within a material, determining the vibrational modes within the crystal lattice. It provides crucial information such as group velocities, which are directly linked to thermal conductivity, and the phonon density of states (DOS), which influences heat capacity. Current methods for calculating phonon dispersion include density functional perturbation theory (DFPT)~\cite{baroni2001phonons} and finite displacement methods.

Several ML models have been developed to predict phonon dispersion relations quickly.
Okabe et al.~\cite{okabe2023virtualnodegraphneural} (Fig.~\ref{Fig_phonon_dynamics} (a)) introduced a virtual node graph neural network (VGNN) to predict $\rm \Gamma$-phonon spectra and full phonon dispersion directly from atomic coordinates. VGNN defines virtual nodes between the connection of each node, which avoid a fixed number of output dimensions.
Fang et al.~\cite{fang2024phonon} (Fig.~\ref{Fig_phonon_dynamics} (b)) presented a E(3)-equivariant graph neural network (GNN) to predict the phonon modes of molecules and crystals. The GNN model learned the underlying potential energy landscape of an atomic structure and calculated its second derivative Hessian matrices to get the harmonic force constants and further predict the phonon dispersions. 
Gurunathan et al.~\cite{gurunathan2023rapid} (Fig.~\ref{Fig_phonon_dynamics} (c)) introduced an Atomistic Line Graph Neural Network (ALIGNN) to predict phonon density of states (DOS) and derived thermodynamic properties. ALIGNN combines atomistic graph representations with line graph connectivity to accurately capture the spectral features of the phonon DOS. Based on the predicted phonon DOS, they can categorize the dynamical stability of materials and derive several thermodynamic properties, including the heat capacity, vibrational entropy, and the isotopic phonon-scattering rate.
Liu et al.\cite{liu2020leverage} (Fig.~\ref{Fig_phonon_dynamics} (d)) demonstrated how transfer learning could enhance phonon dispersion predictions by leveraging electronic properties, which are computationally cheaper to obtain. This strategy improves generalization across diverse material systems and accelerates the prediction process. Together, these advances in ML-driven phonon dispersion prediction open the door to a faster, large-scale screening of thermal and vibrational properties for novel materials.

\subsection{Phonon scattering}

Phonon scattering governs the linewidth of infrared and Raman spectra, and thermal conductivity in most insulators and semiconductors~\cite{ziman2001electrons}. It is very difficult to model due to its dependence on complex anharmonic interactions. Accurate predictions of phonon scattering rates and thermal conductivity have been achieved through first-principles calculations, which rely on the Density Functional Theory (DFT) calculation and solving the Boltzmann transport equation (BTE).
The foundation work for the phonon BTE was laid by Peierls~\citep{peierls1929kinetischen} and was later expanded by Maradudin et al.~\citep{maradudin1962scattering} who developed three-phonon (3ph) scattering theory. 
Subsequent work by Broido et al.~\cite{broido2007intrinsic} combined \textit{ab initio} force constants with these approaches, enabling robust first-principles predictions of thermal conductivity. This integration significantly advanced the understanding of thermal transport~\citep{ esfarjani2011heat, lindsay2018survey, mcgaughey2019phonon}.
More recently, Feng and Ruan developed the formalism and computational method for four-phonon (4ph) scattering, demonstrating its significance across a variety of materials and temperature ranges~\cite{feng2016quantum, feng2017four}. Their predictions for boron arsenide (BAs) were later confirmed by experiments~\citep{kang2018experimental, tian2018unusual, li2018high}. The 4ph scattering has since been shown to play a significant role in the thermal conductivity and thermal radiative properties of numerous materials~\citep{yang2019stronger, kundu2021ultrahigh, xia2020high, guo2023anharmonic, yang2020observation}.

However, the first-principles calculations of phonon-phonon scattering, especially four-phonon scattering, are highly expensive. The high computational cost arises from the need to compute a large number of scattering processes. For 3ph scattering, we need to evaluate each possible triplet of phonon modes, which scales with $N^3$ ($N$ is the number of \textbf{q}-points in the Brillouin zone). For 4ph scattering, the computational cost grows even more steeply, following a scaling of $N^4$. This exponential increase in complexity makes 4ph calculations orders of magnitude more expensive than 3ph processes, especially for materials with complex structures or at high temperatures, where a larger number of phonon modes are thermally activated. 

To address these challenges, several ML methods have been developed.
Guo et al.~\cite{guo2023fast} first introduced a machine learning surrogate model to predict the scattering rates for individual phonon processes (Fig.~\ref{Fig_phonon_dynamics} (e)). By training an MLP on a small, analytically calculated subset of scattering processes, the model can then predict the scattering rates for the remaining large number of processes, bypassing the need for direct calculations. This approach accelerated thermal conductivity predictions by up to 70 times. The use of transfer learning further improved the model's performance. 
Srivastava et al.~\cite{srivastava2024accelerating} developed a random forest model to predict the phonon relaxation time of each phonon mode (Fig.~\ref{Fig_phonon_dynamics} (f)).  By capturing the complex, nonlinear relationships between phonon properties and their relaxation times, the model reduces computational complexity while maintaining high accuracy. Srivastava et al. further combined their approach with Guo et al.'s machine learning framework~\cite{guo2023fast} to create a hybrid framework that achieved a two-order-of-magnitude acceleration in thermal conductivity calculations. These machine-learning-based surrogate strategies significantly reduce computational costs. 

In addition to accelerating the calculation of individual scattering rates, new methods have been proposed to reduce the number of scattering processes that must be explicitly computed. 
Guo et al.\cite{guo2024sampling} presented a method based on statistical sampling and maximum likelihood estimation (MLE) (Fig.~\ref{Fig_phonon_dynamics} (g)). Instead of calculating every possible phonon-phonon interaction, a small random sample of scattering processes is computed, and the total scattering rate is estimated from this subset, leveraging the Central Limit Theorem (CLT). This method achieved acceleration of three to four orders of magnitude compared to traditional rigorous calculation while maintaining a relative error of less than 10\%. 
Given its effectiveness and efficiency, the sampling method has been widely used in the calculation of the thermal conductivities of complex materials~\cite{guo2024first,guo2024first_optics,cui2024effect,wei2024hierarchy,wu2024impact,tang2024effects,alkandari2025anisotropic,guo2025electronic}. Further improvements were made by Zhang et al.~\cite{zhang2024cryogenic} (Fig.~\ref{Fig_phonon_dynamics} (h)), who combined the MLE sampling approach with a phonon frequency cutoff method. At low temperatures, many high-frequency phonon modes are not thermally activated and, therefore, do not contribute significantly to thermal conductivity. By excluding these high-frequency phonons from the scattering calculations, the computational cost is reduced while still preserving high accuracy. This approach is particularly effective for materials at cryogenic temperatures. Besides these works, Gokhale and Jain~\cite{gokhale2025non} proposed a non-uniform Brillouin zone sampling method for studying layered materials, reducing the computational cost by a factor of ten while maintaining relative error within 12\% compared with the uniform grid approach. Malviya and Ravichandran~\cite{malviya2025efficient} present a low-rank spectral method that accelerates the prediction of wave-like heat transport at cryogenic temperatures by over a million times. Guo et al.~\cite{guo2025fourphonongpugpuacceleratedframeworkcalculating} develop a CPU-GPU heterogeneous computing framework to accelerate the phonon scattering calculation by ~25$\times$ without sacrificing accuracy.

\section{ML interatomic potentials}
Molecular Dynamics (MD) simulations are widely used to quantify and understand thermal transport physics at the atomic scale. They have proved advantageous for the discovery and enhancement of electronics, energy storage and conversion applications~\cite{Volz2000, Huang2008,Henry2008,Henry2009,Hu2009,Mcgaughey2009, Feng2017, Feng2019,du2023machine}. Unlike other thermal transport simulation methodologies, MD captures temperature and size-dependent simulations under both equilibrium and non-equilibrium conditions while naturally including higher-order anharmonicity and inelastic scattering. Their physics is essential to investigate heat transfer in bulk materials~\cite{Zhou2015md,feng2017four}, nanostructures, interfaces~\cite{Saskilahti2014,Shi2018, Feng2017, Feng2019, Gordiz2019, zhou2025thermalboundaryconductancestandalone}, amorphous materials~\cite{Lee1991, Zhang2016}, novel two-dimensional (2D) materials like graphene~\cite{Hu2009}, MoS\textsubscript{2} ~\cite{Xu2019}, alloys~\cite{Skye2008, Larkin2013} etc. The Green-Kubo (GK) formalism is commonly used for equilibrium molecular dynamics (EMD), while the non-equilibrium molecular dynamics (NEMD) is used to simulate a heat sink and heat source-based system. Additionally, various formalisms have been developed to decompose MD atomic trajectories for accurate spectral insights of phonon properties~\cite{Saskilahti2014,Zhou2015md,Gordiz2015, Feng2017}. 

MD simulations rely on interatomic potentials to model atomic interactions to perform time-evolving simulations using classical mechanics. Traditional empirical interatomic potentials (EIPs) such as Lennard-Jones potential~\cite{lennard1924determination}, Tersoff potential~\cite{tersoff1988new}, Morse potential~\cite{morse1929} etc. are parametrically fitted mathematical functions representing the potential energy surface (PES) of a material. While efficient, EIPs can struggle with accurately characterizing thermal properties, especially for novel materials with complex crystal structures, interfaces, and nanostructures. On the other hand, \textit{ab initio} molecular dynamics (AIMD) simulations based on quantum mechanical principles~\cite{marx2000ab,iftimie2005ab} offer very high accuracy with a significantly higher computational cost. Typically accurate thermal property evaluation requires extended MD simulations (1–10 ns) with timesteps on the order of 0.1-1 fs to resolve vibrational modes. Furthermore, characterizing complex nanostructures, interfaces, etc., requires a larger simulation domain to mitigate the limitations of size effects, such as a limited phonon mean-free path. Such large-scale AIMD simulations are impractical due to computational limitations. 

Machine learning interatomic potentials (MLIPs) have emerged as a potential solution to bridge the gap between computationally expensive AIMD and parametrically limited EIPs. MLIPs are trained on small-scale high-fidelity datasets from static DFT calculations and finite temperature AIMD simulations, as shown in Fig.~\ref{Fig_MLIP}(a). They offer a faster and more accurate alternative to characterize thermal properties with near-first-principles accuracy. Behler and Parinello first demonstrated the application of neural networks to describe the potential energy surface of bulk silicon in 2007~\cite{Behler2007}. Subsequently, various ML models have been employed for MLIPs such as the neural network potential (NNP)~\cite{Wang2018deepmd, Zeng2023deepmd, Lee2019, Huang2019}, Gaussian approximation potential (GAP)~\cite{Bartok2010gap, Bartok2015}, moment tensor potential (MTP)~\cite{Shapeev2016}, spectral neighbor analysis method (SNAP)~\cite{Thompson2015}, atomic cluster expansion (ACE)~\cite{Drautz2019}, among others~\cite{Barry2020, Wyant2021}. Models like SNAP and MTP use linear functions for the descriptors, which creates the need for more complex features for complex material systems. Neural networks can capture the non-linear relations more effectively at the cost of computational efficiency. GAPs are non-parametric models as they adapt during the training process. However, for most ML models, the computational cost scales up as the complexity increases, limiting MD simulations with larger system sizes and longer run times.

MLIPs have also emerged as efficient surrogate models for DFT calculations to estimate interatomic force constants (IFCs) essential for BTE-solvers that estimate thermal properties~\cite{Mortazavi2021}. The BTE solution is capable of capturing both harmonic and anharmonic effects. The anharmonic IFCs are often solved using the finite-displacement method for supercells with specific atoms displaced from the equilibrium position. The number of displaced structures and corresponding DFT calculations increases significantly for higher-order terms. Furthermore, the complexity of crystals and their lack of symmetry can exacerbate the need for more sampling. MLIPs can reduce the computational time and power required to evaluate each of the displaced structures to a few seconds. Various simpler and less complex models like least absolute shrinkage and selection operator (LASSO), singular-value decomposition (SVD), etc., have proved useful for capturing temperature-dependent effects on IFCs~\cite{Eriksson2019,Knoop2024}. However, their application is limited to obtaining IFCs and are not suitable for MD simulations. In the following sections, we focus on full-scale MLIPs capable of both accurate MD simulations and obtaining IFCs. 

\subsection{Bulk Materials}
\begin{figure}[h]%
\centering
\includegraphics[width=1.0\textwidth]{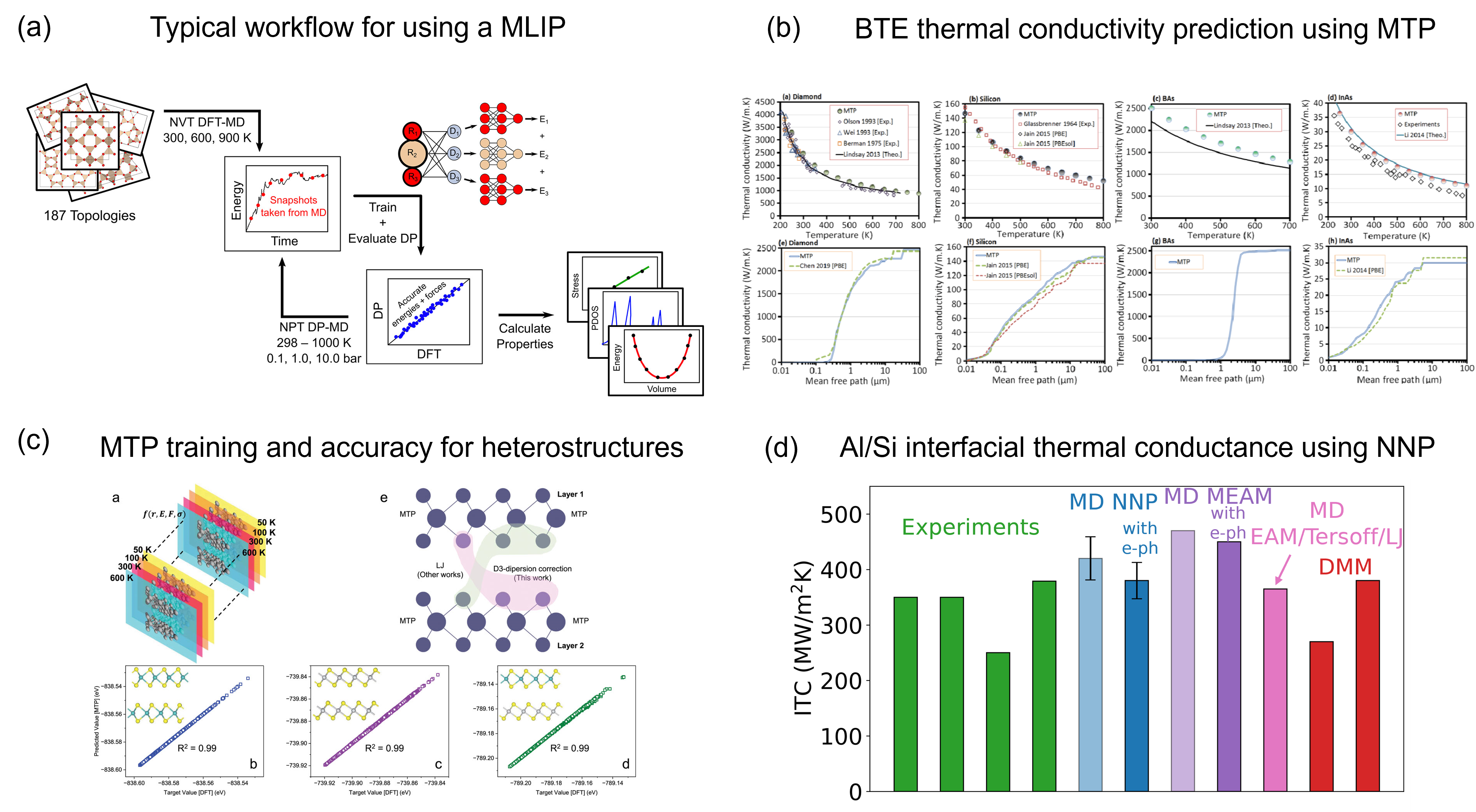} 
\caption{\textbf{MLIP-driven predictions of thermal properties.} (a) Workflow of generating a dataset using \textit{ab initio} molecular dynamics simulations and training an MLIP model for property prediction, as adapted from Sours and Kulkarni~\cite{Sours2023}. (b) Thermal conductivity predictions for Diamond, Silicon, BAs, and InAs using MTP/ShengBTE approach, as adapted from Mortazavi et al.~\cite{Mortazavi2021}. (c) Illustration of training dataset for bilayer heterostructures, and the prediction accuracy of MTP for TiS\textsubscript{2}/MoS\textsubscript{2} systems, as adapted from Nair et al.~\cite{Nair2024}. (d) Interfacial thermal conductance estimate from NNP-driven NEMD simulations compared to experiments and other simulation techniques, as adapted from Khot et al.~\cite{Khot2025}.
}\label{Fig_MLIP}
\end{figure}

Many MLIPs have been developed and used for detailed investigation of thermal properties of bulk materials~\cite{Korotaev2019,Qian2019,Korotaev2020,Mangold2020,Li2020ga2o3,Li2020,Dai2020,Liu2021,Wang2021,Zeng2021,Huang2021,Han2021,Dai2021,Verdi2021,Ouyang2022,Zeng2021BaAg2Te2,Choi2022,Zhang2022bite,Ouyang2022SnSe,Sun2022,Tang2023bas,Tang2023bn,Tiwari2024}. Qian et al. developed a GAP, and Li et al. used an NNP to predict the thermal conductivity of silicon in the crystalline and amorphous phases~\cite{Qian2019, Li2020}. Korotaev et al. demonstrated the utility of MTP
for complex compounds like CoSb\textsubscript{3}~\cite{Korotaev2019}. They demonstrated a computational acceleration of 80,000$\times$ for a 128-atom system with MTP-based prediction compared to DFT calculations. Mortazavi et al. developed the MTP/BTE extension for thermal conductivity predictions and demonstrated its accuracy and efficiency for semiconductors with narrow to ultrawide bandgaps, as shown in Fig.~\ref{Fig_MLIP}(b)~\cite{Mortazavi2021}. The MLIP approach has proved beneficial for the extensive study of BAs, a high thermal conductivity, wide bandgap semiconductor~\cite{Liu2021,Mortazavi2021,Ouyang2022,Tang2023bas}.  Liu et al. used an MTP to calculate the thermal conductivity of c-BAs and w-BAs at the three-phonon and four-phonon levels, which were within 8\% of the DFT results for the whole temperature range~\cite{Liu2021}. The w-BAs require 2624 calculations to obtain IFCs up to the fourth order, which accounts for about 1600 hours of computational time using 2 nodes with 40 CPU cores each. In comparison, the MTP utilized $\sim$230 hours of compute time for the AIMD dataset, and $\sim$10 hours for the training process. Various MLIPs have been developed for thermoelectrics and their thermal properties for materials like CoSb\textsubscript{3}, SnSe, Sb\textsubscript{2}Te\textsubscript{3}, Bi\textsubscript{2}Te\textsubscript{3}, Tl\textsubscript{3}VSe\textsubscript{4}, BaAg\textsubscript{2}Te\textsubscript{2}~\cite{Korotaev2019,Ouyang2022SnSe,Sun2022,Huang2021,Zhang2022bite, Zhang2023super}. Similar MLIPs have been used to study the phonon thermal conductivities for other classes of materials like ceramics~\cite{Verdi2021, Tiwari2024}, skutterudites~\cite{Korotaev2019,Korotaev2020}, perovskites~\cite{Wang2021}, intermetallics~\cite{Mangold2020}, and high entropy alloys~\cite{Dai2020,Dai2021}.

Besides near-first-principles accuracy, the MLIP acceleration to obtain anharmonic IFCs has enabled in-depth studies for systems at different temperatures and pressure conditions with various stoichiometries, phases, etc. Li et al. showed good agreement of Si thermal conductivity during phase transition at the melting point using the NNP~\cite{Li2020}. Huang et al. and Ouyang etl al. reproduced the phase transition in SnSe in their MTP-based GK EMD, and captured the corresponding temperature and pressure-dependent thermal conductivities~\cite{Huang2021, Ouyang2022SnSe}. Tang et al. investigated the competition between four-phonon scattering and phonon-vacancy scattering in c-BAs, along with their dependence on temperature and vacancy concentration~\cite{Tang2023bas}. Tiwari et al. performed a comprehensive study on Al\textsubscript{2}O\textsubscript{3} to track the thermal conductivity changes from 300 K to 2200 K~\cite{Tiwari2024}. Their work includes the contributions from phonon ($\kappa_{ph}$), diffusion ($\kappa_{diff}$) and radiation ($\kappa_{rad}$), which are essential at ultrahigh temperatures.

\subsection{Two-Dimensional Materials}

Beyond bulk three-dimensional materials, MLIPs have been used to study thermal properties of novel 2D materials~\cite{Gu2019,Zhang2019,Mortazavi2020twoD,Mortazavi2020graphene,Ghosal2021,Zhang2020hbn,Arabha2021,Mortazavi2022,Chen2024,Liu2024}. Gu and Zhao demonstrated the use of SNAP for MoS\textsubscript{2(1-x)}Se\textsubscript{2x} alloys using both EMD and BTE approach~\cite{Gu2019}. Zhang and Sun used the sinusoidal approach to equilibrium molecular dynamics (SAEMD) and time domain normal mode analysis (TDNMA) for silicene using GAP~\cite{Zhang2019}. Their work demonstrates that GAP-based MD outperforms BTE, as MD captures large random perturbations of certain Si atoms, which move in and out of the 2D plane of the material. Mortazavi et al. have studied various 2D materials like graphene, MoS\textsubscript{2}, carbon nitrides, borophene etc~\cite{Mortazavi2020twoD,Mortazavi2020graphene,Mortazavi2022}. Recently, the utility of MLIPs has been demonstrated for thermal investigations of 2D heterostructures like graphene-borophene~\cite{Mortazavi2020graphene,Chen2024}, TiS\textsubscript{2}/MoS\textsubscript{2}~\cite{Nair2024}, and MoS\textsubscript{2}-WS\textsubscript{2}~\cite{Liu2024}. Figure~\ref{Fig_MLIP}(c) by Nair et al. shows the accuracy of the MTP trained for TiS\textsubscript{2}/MoS\textsubscript{2} bilayer heterostructures while effectively capturing the short-range van der Waal's corrections~\cite{Nair2024}. Their NEMD simulations showed that these bilayer heterostructures possess significantly higher thermal conductivity compared to graphite used for battery energy storage.

\subsection{Interfaces}

Interfacial thermal resistance (ITR) poses a critical challenge for the current and next-generation semiconductor devices. As the device feature size reduces, the number of interfaces and power density increase. Hence, it is crucial to understand the physical mechanisms and the solutions to minimize or maximize the ITR. However, precisely characterizing the nanoscale interfacial thermal transport is exceptionally challenging. Furthermore, the ITR is significantly influenced by complex mechanisms like inelastic scattering, phonon local nonequilibrium, and interfacial phonon modes~\cite{Feng2017,Feng2019,Gordiz2019,Cheng2021,Khot2025}. NEMD simulations offer a pathway to study interfacial thermal transport while capturing these physical mechanisms. However, their results are sensitive to interfacial interactions. Typically, the interfacial atomic interactions are either described using approximations like arithmetic and geometric mean or by fitting a simple Lennard-Jones model. These EIP-driven approaches to model the interface can compromise the primary objective of interfacial thermal transport investigations using NEMD simulations.

Training MLIPs using interfacial supercells at the \textit{ab initio} level has enabled tackling the challenge of capturing interfacial interactions in various semiconductor/semiconductor~\cite{Wyant2021, Cheng2021,Zhang2023super,Nair2024} and metal/semiconductor interfaces~\cite{Adnan2024,Khot2025} recently. Wyant et al. combined the SNAP with a translationally invariant Taylor expansion to study Ge/GaAs interface~\cite{Wyant2021}. Chen et al. used the NNP NEMD approach to compare Si/Ge TBC with their experimental measurements~\cite{Cheng2021}. NNP, neuroevolution potentials (NEP) and MTP were used to obtain the thermal conductivities of heterostructures of Sb\textsubscript{2}Te\textsubscript{3}/Bi\textsubscript{2}Te\textsubscript{3}, GeTe/Sb\textsubscript{2}Te\textsubscript{3} and TiS\textsubscript{2}/MoS\textsubscript{2}, respectively~\cite{Zhang2023super,Nair2024,Wang2025GeTe}. Diamond is considered to be a future UWBG semiconductor~\cite{Wong2021uwbg}, however, its ITR with metal contacts is little known. Adnan et al. used the MTP to model metal/diamond interfaces for candidate metals like Al, Mo, Zr, and Au~\cite{Adnan2024}. Khot et al. recently developed a NNP trained on both bulk and interface supercells for the Al/Si interface~\cite{Khot2025}. As shown in Fig.~\ref{Fig_MLIP}(d), their ITC estimates using NEMD simulations are within 8\% of the experimental consensus achieved in the last decade. They use spectral analysis to demonstrate the interfacial phonon modes and phonon local non-equilibrium at the interface with the near-first-principles accuracy of the NNP-NEMD simulations. 

These studies have collectively demonstrated the utility of various ML architectures for obtaining PES and performing molecular simulations with near-first-principles accuracy. Although MLIPs have proven clear advantages over classical EIPs, the prediction errors and overfitting of ML models can lead to systematic discrepancies in property predictions. This issue can be further exacerbated, particularly for the lattice thermal conductivity (LTC), which depends on the accurate prediction of interatomic forces. Various MLIP-driven studies of high thermal conductivity materials like CoSb\textsubscript{3}~\cite{Korotaev2019}, Si~\cite{Qian2019}, GaAs~\cite{Wu2024}, and graphene~\cite{Wu2024} have shown systematic underprediction of the LTCs compared to experimental benchmarks. Wu et al. performed a systematic study and concluded that the force prediction error is the primary reason for the LTC underprediction~\cite{Wu2024}. They artificially introduced force errors in NEMD simulations at various levels and extrapolated the LTC at the limit of zero force error for c-Si, GaAs, graphene, and PbTe. The resulting LTC values were found to be in closer agreement with experimental results over a broad temperature range. Recently, Zhou et al. built on this previous work and interpreted this LTC underestimation as a ’pseudo-isotope effect’ which results in slightly higher phonon scattering~\cite{Zhou2025}. A second-order force correction term was introduced in their work to improve the robustness and accuracy of LTC predictions. Further work is needed to minimize such artifacts from MLIP predictions and to develop more optimized architectures that promote faster and more reliable predictions.

\section{ML for solving radiative heat transfer}

\begin{figure}[h]%
\centering
\includegraphics[width=1.0\textwidth]{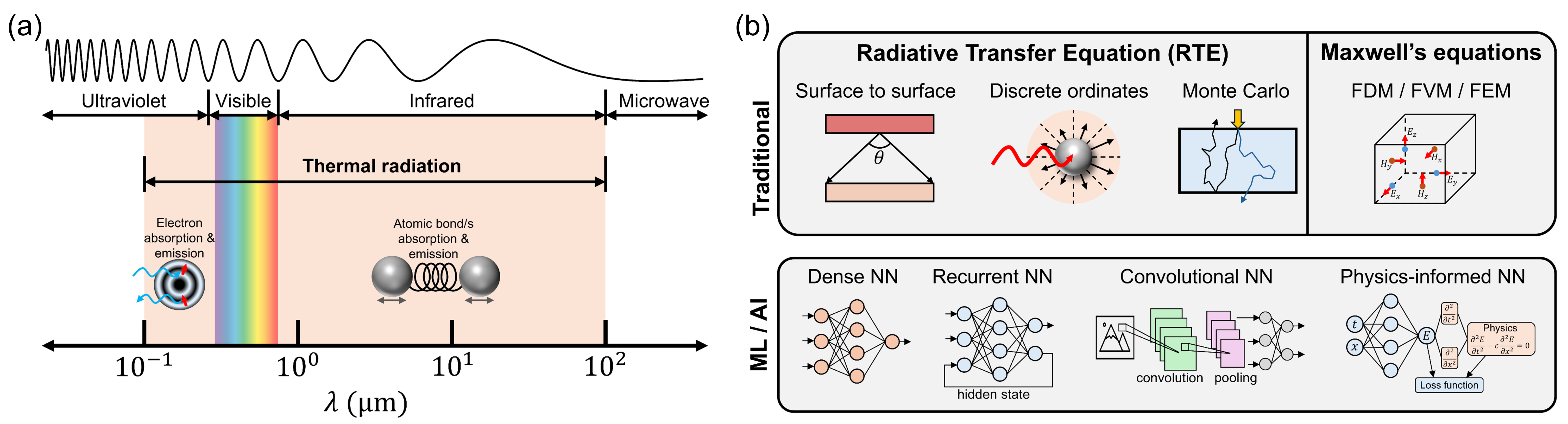} 
\caption{(a) Electromagnetic spectrum from ultraviolet through microwave, highlighting the band typically considered as thermal radiation. (b) Traditional and ML/AI methods used to solve radiative heat transfer.
}\label{Fig_Carne1}
\end{figure}

Radiative heat transfer refers to heat transfer due to the absorption and emission of electromagnetic waves, also known as photons, which can occur across a broad wavelength spectrum. Typically, thermal radiation considers the ultraviolet, visible, and infrared bands from 0.1-micron to 100-micron wavelength as highlighted in Fig.~\ref{Fig_Carne1}(a))~\cite{FundHeatTransf}. There are several fundamental mechanisms inducing photon absorption and emission. For thermal radiation, this is commonly due to electronic transitions at shorter wavelengths, and vibrational and rotational transitions in atomic bonds at longer wavelengths~\cite{Modest}. Thermal radiation has significant impacts across many fields and applications. Solar applications, including solar power~\cite{Zhao2011_solarpower}, solar heating~\cite{Lee2012_solarheating}, radiative cooling~\cite{li2021ultrawhite}, and climate modeling~\cite{Giorgi1993_climate}, clearly rely on solutions to radiative heat transfer to model solar irradiation. Combustion applications, including furnaces~\cite{Hottel1970_furnace} and gas turbines~\cite{Jones2005_turbine}, require radiative heat transfer simulations due to the high temperature (radiative heat transfer scales with temperature to the fourth power) as well as the absorption and scattering induced by the soot and particulates within the system. Space systems necessitate radiative heat transfer simulations due to the near-zero conduction in space, meaning radiation is the dominate heat transfer mechanism~\cite{Tachikawa_space}, and to understand the impacts of Martian and Lunar regolith coatings on radiators and spacecraft~\cite{Cannon2022_lunar}. Laser systems, such as for advanced manufacturing~\cite{Cook2020_manufacturing} and directed energy weapons~\cite{Perram2004_laser}, utilize radiative transfer simulations to understand heat load and safety under varying environmental conditions. Fast, efficient, and reliable radiative heat transfer simulation and modeling are critical to ensuring the advancement and development of these applications.

We first overview the mainstream simulation methods for thermal radiation. Since radiative heat transfer arises from electromagnetic waves, thermal radiation can be modeled with Maxwell’s equations~\cite{Modest}. For simple geometries, analytical solutions exist, which provide fast and accurate solutions. Common analytical solutions include Mie theory~\cite{MieTheory}, which describes incident radiation on a homogenous sphere, and the transfer matrix method~\cite{PrincOptics}, which can solve light propagation through multi-layer plane parallel media. To model complex geometries, numerical methods are required to solve Maxwell’s equations such as the finite difference method (FDM), finite volume method (FVM), or finite element method (FEM) either in the time or frequency domain~\cite{FEMmethods, FDTDmaxwell}. However, these methods are tremendously computationally expensive for complex geometries due to strict meshing requirements. An alternative method for simulating radiative transfer is through the Radiative Transfer equation (RTE), which is the typical approach for solving radiative heat transfer problems~\cite{Modest}. The RTE treats light as incoherent, meaning it does not capture the wave effect of light, and accounts for absorption, emission, and scattering or radiation. Solving the RTE is computationally advantageous for macro-scale geometries where the wavelength is significantly smaller than the geometric feature sizes, or when the optical properties of small features (e.g. nanoparticles) can be determined through experiment or by simulating Maxwell’s equations on the individual feature. Several analytical solutions of the RTE exist, such as Beer-Lambert’s law for transmission through homogeneous absorbing media~\cite{Modest}. For complex geometries, several numerical methods exist, including surface to surface, discrete ordinates, spherical harmonics, and Monte Carlo simulation~\cite{Modest}. Surface-to-surface models are commonly applied when the medium between each surface is not optically active, as it does not absorb, emit, or scatter light. When the medium is optically active, such as CO\textsubscript{2} gas or combustion soot in air, a more complex model is required to account for absorption, emission, and scattering of light by the medium. Discrete ordinates method, for example, discretizes the angular domain as well as the spatial domain in conjunction with a method like the finite volume method to model radiative transfer between surfaces and within the optically active medium. While this method is fast compared to other radiative transfer simulation methods, it is more computationally expensive than other finite volume methods utilized in computational fluid dynamics (energy, momentum, etc.)~\cite{Viskanta2008}. This is due to the angular discretizations requiring the discretized equation to be solved multiple times at each cell. Alternatively, Monte Carlo simulations are also commonly used to simulate radiative transfer within optically active media~\cite{Howell1964, Howell1998}. Monte Carlo simulations, which stochastically model a large number of individual energy bundles, provide several important benefits over other methods, including high accuracy, the ability to quantify error, and efficient parallelization. However, Monte Carlo simulations are often considerably more computationally expensive than methods such as discrete ordinates~\cite{Howell1998}.
Based on our discussion above, a common limitation of these simulation methods is the computational cost for accurate solutions.

\begin{figure}[h]%
\centering
\includegraphics[width=0.8\textwidth]{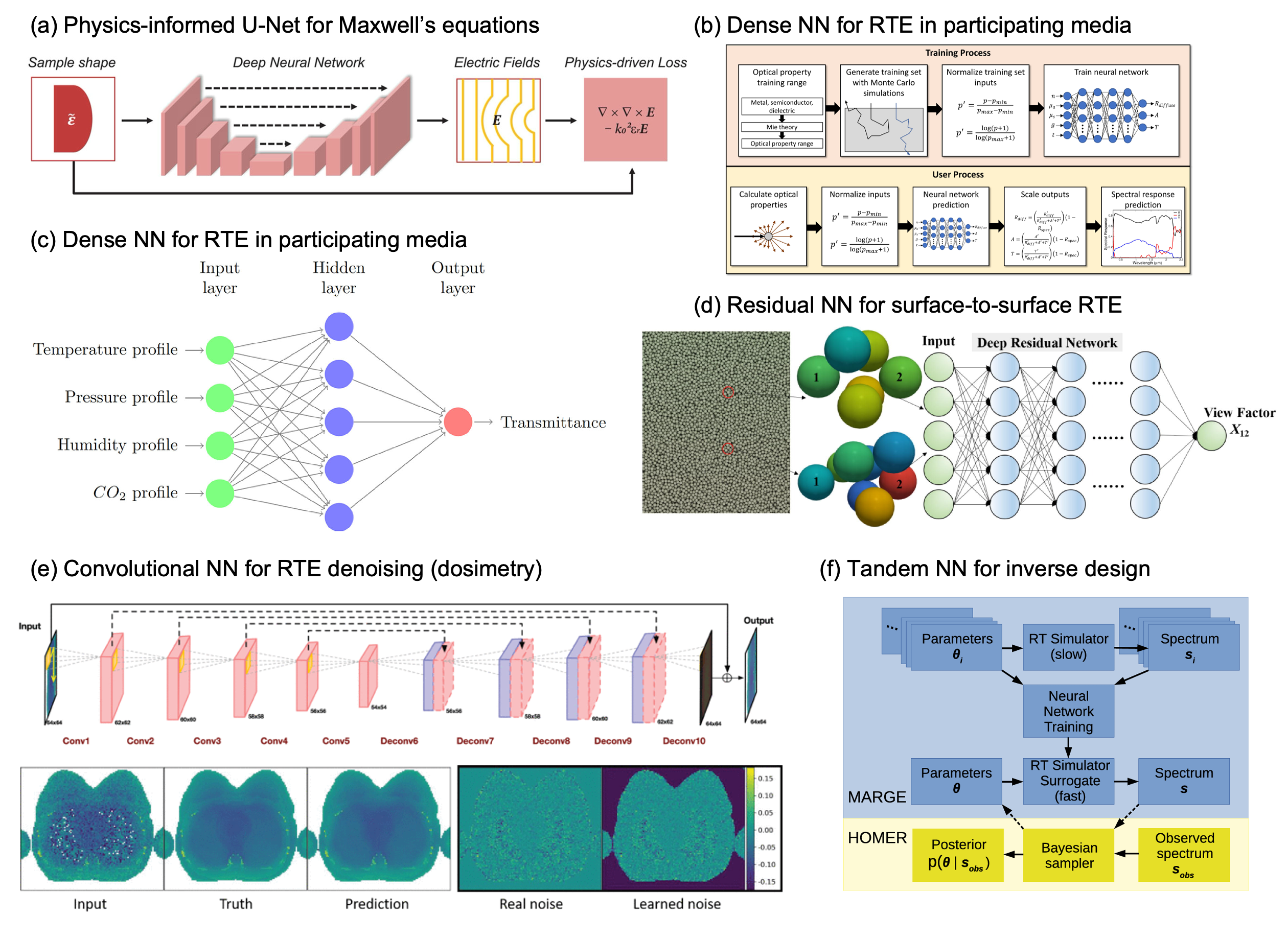} 
\caption{\textbf{ML prediction of radiative heat transfer.} (a) Physics-informed U-net for solving Maxwell's equations, as adapted from Lim et al.~\cite{Lim2022}, (b) Dense NN for solving radiative transfer in participating media, as adapted from Carne et al.~\cite{carne2023accelerated}, (c) Dense NN for solving radiative transfer in participating media, as adapted from Stegmann et al.~\cite{Stegmann2022}, (d) Residual NN for solving surface-to-surface radiative transfer, as adapted from Wu et al.~\cite{Wu2022}, (e) CNN for dosimetry denoising, as adapted from Peng et al.~\cite{peng2019cnn}, (f) Tandem NN for inverse design of colored radiative cooling films, as adapted from Himes et al.~\cite{Himes2022}.
}\label{Fig_Carne2}
\end{figure}

Various AI and ML methods have been implemented to overcome these computational power limitations, including dense, recurrent, convolutional, and physics-informed neural networks Fig.~\ref{Fig_Carne1}(b)). Physics-Informed Neural Networks (PINNs) have become a popular tool for accelerating physics simulations~\cite{zhou2023physics,zhou2023physicsAHT,li2025physics,zheng2024hompinns}. PINNs have a unique advantage over other ML models in that they do not require training data generated by running other numerical simulations, where additional error could be introduced. Instead, since neural networks are readily differentiable, the loss function can be the residual of Maxwell’s equations. For example, Zhang et al.~\cite{Zhang2021} trained a dense PINN to predict the electric and magnetic fields where the loss function considers the initial conditions, boundary conditions, and Maxwell’s equations simultaneously. Additionally, Lim et al.~\cite{Lim2022} (Fig.~\ref{Fig_Carne2}(a)) trained MaxwellNet, a physics-informed U-net to predict the electric field through micro-lenses. While using PINNs to solve Maxwell’s equations provides many advantages, they also have disadvantages, such as the poor generalizability to different geometries.

To accelerate solving RTE, several ML models have been developed. For example, Carne et al.~\cite{carne2023accelerated} (Fig.~\ref{Fig_Carne2}(b)) used a dense neural network to predict the spectral response through plane-parallel nanoparticulate media. Furthering this work, they developed an RNN to predict the spectral response of multi-layer plane-parallel media~\cite{carne2025overcomingcursedimensionalityenabling}. Stegmann et al.~\cite{Stegmann2022} (Fig.~\ref{Fig_Carne2}(c)) used a dense neural network to predict atmospheric transmittance based on the temperature, pressure, humidity, and CO\textsubscript{2} profile. Furthermore, Kearney et al.~\cite{Kearney2018} developed DoseNet, a convolutional neural network (CNN) to predict radiative transfer for dosimetry absorption maps. Each of these studies uses ML to completely replace the radiative transfer model it is trained on. Alternatively, there are ML models that combine with traditional radiative transfer solvers to provide an accelerated solution. Wu et al.~\cite{Wu2022} (Fig.~\ref{Fig_Carne2}(d)) used a residual neural network to predict view factors in dense granular systems. View factor calculations are a significant portion of the computational expense in surface-to-surface radiative transfer models, which are then used to calculate the radiative heat transfer. Additionally, Peng et al.~\cite{peng2019cnn} (Fig.~\ref{Fig_Carne2}(e)) developed MCDNet, a CNN for denoising dosimetry absorption maps. This network takes in a low-resolution Monte Carlo solution and increases accuracy through denoising, allowing for 76-fold speedups over an equivalent high-resolution Monte Carlo simulation.


So far, we have mainly discussed the ``forward" problem,  where the geometry and material properties are given to solve the radiative heat transfer. However, the inverse problem is regularly required where the radiative transfer is known and either properties or geometric features are solved for, such as the atmospheric parameters~\cite{Chahine1970} or tissue properties~\cite{Klose2010}. 
A common solution technique for the inverse problem is to pair a traditional forward solver with an optimization algorithm. The material property or geometric feature is optimized until the predicted radiative transfer closely matches the known solution. Due to this being an iterative process, the inverse problem is considerably more computationally expensive than the forward problem, making it an attractive target for machine learning acceleration. For example, Kim et al.~\cite{Kim2024_inverse} trained a tandem neural network to efficiently solve the inverse problem to design nanoparticle embedded radiative cooling films. A tandem neural network trains both the forward and inverse radiative transfer processes simultaneously based on training data from the forward process, providing significant time savings compared to traditional methods. Furthermore, Himes et al.~\cite{Himes2022} (Fig.~\ref{Fig_Carne2}(f)) used a CNN to accelerate the inverse process of determining atmospheric properties of exoplanets based on a measured spectrum, providing a 9-fold speedup compared to traditional methods.

\section{ML-assisted design of thermal radiative energy devices}

In the last section, we discussed using ML to directly solve the governing equations of radiative heat transfer. In this section, we discuss how to leverage those surrogate models to enable inverse design and optimization of functional thermal radiative devices.
The design of thermal radiative energy devices is a rapidly growing field with critical applications in radiative cooling~\cite{felicelli2022thin,li2021ultrawhite,raman2014nature,kim2024review}, thermophotovoltaics~\cite{daneshvar2015thermophotovoltaics}, thermal cloaking~\cite{zhu2020high,han2014full}, imaging~\cite{he2021infrared,chen2023meta}, and energy harvesting~\cite{wang2015highly,paul2022nano,tian2013review}. These devices rely on precise control of thermal radiation, which can be achieved by tailoring their structure and material properties to emit, absorb, or reflect specific wavelengths of thermal radiation. However, the design of such devices often relies heavily on human intuition, trial-and-error experimentation, or exhaustive parameter sweeps, all of which are computationally expensive and time-consuming. As the complexity of device design increases, these conventional approaches struggle to explore the vast design space effectively.
Machine learning has emerged as a powerful tool for efficiently navigating such large design spaces, which would reduce computational costs and discover novel device architectures that would be difficult to identify using traditional methods~\cite{chen2025metamatbench}.  ML-assisted design of thermal radiative energy devices generally falls into two main categories: optimization-based design and generative design approach. Both approaches leverage advanced ML algorithms to reduce computational complexity and improve design precision.

\begin{figure}[h]%
\centering
\includegraphics[width=0.6\textwidth]{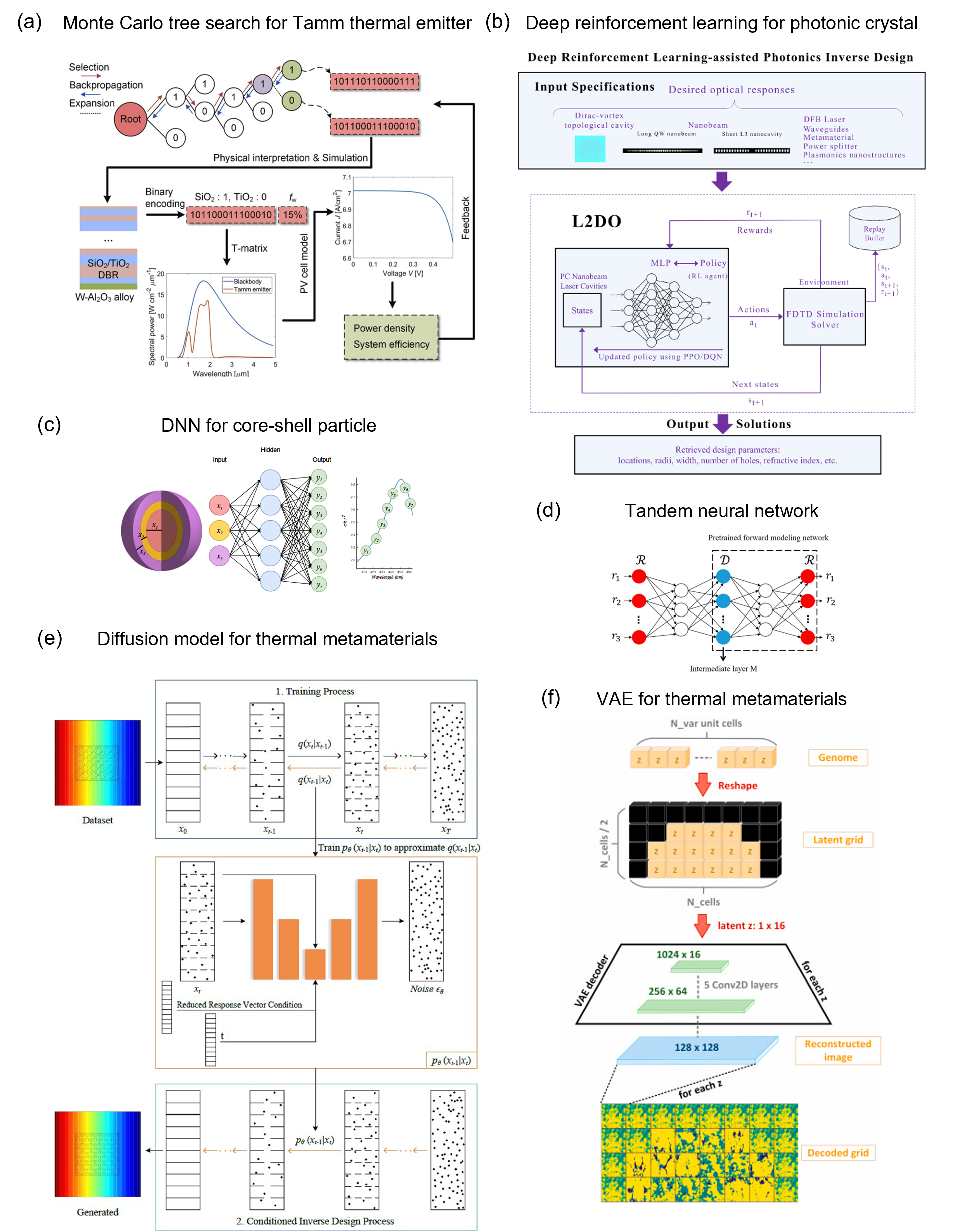} 
\caption{\textbf{ML-assisted design of thermal radiative energy devices.} (a) Machine-learning Monte Carlo tree search for Tamm thermal emitter in TPV systems, as adapted from Hu et al.~\cite{HU2020Tamm}, (b) Deep reinforcement learning-based inverse design of photonic crystals, as adapted from Li et al.~\cite{li2023deep}, (c) Fully connected neural network for designing core-shell particle, as adapted from Peurifoy et al.~\cite{peurifoy2018nanophotonic}, (d) A tandem neural network for the inverse design of multilayer photonic devices, as adapted from Liu et al.~\citep{liu2018training}, (e) Diffusion model for the generative design of thermal metamaterials, as adapted from Liu et al.~\cite{liu2024diffusion}, (f) VAE model for the generative design of thermal metamaterials, as adapted from Ignuta-Ciuncanu et al.~\cite{ignuta2024generative}.
}\label{Fig_inverse}
\end{figure}

\subsection{Optimization-based approach}

In the optimization-based approach, the design of thermal radiative devices is formulated as an optimization task where the device's structure, material composition, and geometric features are the design variables. The objective of this approach is to minimize the difference between the actual radiative properties of the device and the ideal target properties. These properties typically include emissivity, absorptivity, and reflectivity at specific wavelengths or across specific spectral ranges. 
Optimization techniques such as Bayesian optimization~\cite{sakurai2019ultranarrow,zhang2025active}, genetic algorithms~\cite{carne2024true,xu2024quantum}, and quantum annealing~\cite{kim2022high,xu2025quantum} have proven to be highly effective for this purpose.
Given that physics-based simulations of thermal radiation, such as finite-difference time-domain (FDTD) and FEM, are computationally expensive, ML surrogates are frequently employed. These surrogate models approximate the relationship between device configurations and their radiative properties. Once trained, they can predict the properties of new configurations much faster than rigorous simulations. 
For example, 
Hu et al.~\cite{HU2020Tamm} developed a machine-learning-based Monte Carlo tree search algorithm to optimize a Tamm emitter, targeting improved power density and system efficiency of the TPV system (Fig.~\ref{Fig_inverse}(a)). Similar work has also been performed by Bohm et al.~\cite{Zhang2024} for a tungsten emitter using a deep learning model to save the computational cost of Rigorous coupled-wave analysis (RCWA) calculation.
Li et al.~\cite{li2023deep} developed a deep reinforcement learning-based inverse design framework for photonic crystal design for nanoscale laser cavities (Fig.~\ref{Fig_inverse}(b)). 
Carne et al.~\cite{carne2024true} developed a BaSO4-based radiative cooling paint with maximized solar reflectivity. By training a neural network-based surrogate model to predict radiative properties~\cite{carne2023accelerated,carne2025fos}, they replaced complex Monte Carlo simulations, leading to significant computational speed-ups of the evolutionary algorithm optimization process.

\subsection{Generative model approach}

Generative design frameworks take a different approach, aiming to directly produce configurations of thermal radiative devices that meet specific design criteria. Typical generative machine learning models include autoencoder~\cite{Bank2023}, generative adversarial networks (GANs)~\cite{goodfellow2014generative}, variational autoencoders (VAEs)~\cite{kingma2013auto,rezende2014stochastic,chen20243d}, and diffusion models~\cite{song2020score,song2019generative,sohl2015deep,ho2020denoising}. Unlike optimization methods, which search for optimal configurations from a predefined design space, generative models directly generate device structures with desired properties, bypassing the need for an explicit search process. For example,
Peurifoy et al.~\cite{peurifoy2018nanophotonic} developed a fully connected NN for designing core-shell particles with target wavelength-dependent scattering properties (Fig.~\ref{Fig_inverse}(c)). The input is set as the diameter and thickness of core-shell particles, and the output is set as the scattering cross-section at discrete wavelengths. The result shows a good reproduction of the desired properties. 
Liu et al.~\citep{liu2018training} presented a tandem neural network to design a multilayer thin film composed of $\rm SiO_2$ and $\rm Si_3N_4$ to obtain the target transmission spectrum (Fig.~\ref{Fig_inverse}(d)). They claimed that combining forward modeling and inverse design in a tandem architecture could overcome data inconsistency issues, thereby accelerating the training speed of deep neural networks. Guan et al.~\cite{guan2023machine} also applied a tandem neural network for the inverse design of radiative cooling material.
The structure enables high solar transmittance, strong mid-infrared emissivity, and customizable visible colors.
Liu et al.~\cite{liu2024diffusion} developed a diffusion model for the inverse design of thermal metamaterials to enhance thermal transparency (Fig~\ref{Fig_inverse}(e)). 
Ignuta-Ciuncanu et al.~\cite{ignuta2024generative} developed a variational autoencoder model for the design of macroscopic thermal metamaterials. A genetic optimizer is used to explore the latent design space to achieve the temperature and heat flux design goals (Fig~\ref{Fig_inverse}(f)). 
Garcia-Esteban et al.~\cite{García-Esteban2024} employ conditional Wasserstein GANs (CWGANs) to generate synthetic spectral data for near-field radiative systems, enabling improved modeling accuracy in low-data regimes.

\section{Challenges and future directions}

Despite remarkable progress in applying AI to nanoscale heat conduction and radiation, several challenges and limitations remain.
A key bottleneck is the lack of high-quality, standardized datasets. Unlike fields such as computer vision or natural language processing (NLP), thermal sciences lack large-scale annotated datasets necessary for training robust supervised learning models. Encouragingly, recent efforts are helping to bridge this gap. For example, databases of phonon band structures~\cite{phonondb}, anharmonic phonon properties~\cite{ohnishi2025database} and spectral radiative properties~\cite{polyanskiy2024refractiveindex}, along with large-scale initiatives such as the Materials Project~\cite{jain2013commentary}, JARVIS~\cite{choudhary2020jarvis}, OQMD~\cite{kirklin2015open}, and AFLOW~\cite{curtarolo2012aflow}, are increasingly becoming valuable resources for AI model training and validation. Moving forward, the development of benchmark datasets for properties like thermal conductivity and refractive and extinction coefficients will be critical for enabling systematic model comparison and accelerating algorithmic innovation.
Besides, techniques that can learn effectively from limited data are gaining importance. Multi-fidelity modeling~\cite{liu2022leveraging,forrester2007multi} is one such approach. \textcolor{black}{The core principle is to strategically combine datasets from sources of varying accuracy and computational cost. The model is trained primarily on a large volume of ``low-fidelity" data, which is computationally cheap to generate (e.g., from empirical potentials, simplified physical models, or less converged first-principles calculations). This large dataset allows the model to learn the broad trends and fundamental relationships within the design space. Subsequently, a much smaller and more precious set of ``high-fidelity" data, derived from highly accurate simulations and experiments, is used to refine, correct, and calibrate the model. By learning the discrepancy between the low- and high-fidelity predictions, the final model can achieve accuracy approaching that of the high-fidelity method with a limited amount of data. }

Another challenge is the generalization capabilities of current models. Many AI models are trained on narrow domains with specific materials, geometries, or thermal conditions, which limits their applicability to new settings. Enhancing model generality and transferability is therefore a key research priority. For example, universal machine-learned interatomic potentials are being developed to capture diverse behaviors across chemical compositions and structures. 
The idea of foundational models is also gaining interest in the materials community. Similar to large pre-trained models in natural language processing, these models aim to learn general representations of material structures, which can then be fine-tuned for specific tasks such as thermal conductivity or radiative property prediction~\cite{brazil2024mamba,zeni2023mattergen,yang2024mattersim,kishimoto2023mhg}. There has already been work on using these models for finding high thermal conductivity materials~\cite{li2025probing}. 

For real-world AI deployment, particularly in safety-critical or high-precision thermal applications, rigorous uncertainty quantification (UQ) is essential. 
\textcolor{black}{Instead of a single point-value prediction, UQ methods yield a predictive distribution, effectively placing ``error bars" on the output. This is vital for understanding how much trust should be placed in its predictions for engineering design and risk analysis.}
Gaussian processes (GPs) offer a principled way to quantify predictive uncertainty~\cite{gramacy2020surrogates,williams2006gaussian}, though they often do not scale well to large or high-dimensional datasets. As alternatives, deep learning approaches such as Bayesian neural networks~\cite{li2023bayesian, koenig2024uncertain, chen2024fast} and dropout-based techniques~\cite{angelikopoulos2012bayesian} have been adopted for scalable uncertainty estimation.
Interpretability is another crucial issue. While deep neural networks have demonstrated powerful predictive capabilities, they are often criticized for their lack of transparency. Improving model interpretability is not only important for trust and verification but can also lead to new scientific insights~\cite{wang2025machine,feng2025contextualizing}. \textcolor{black}{Techniques such as symbolic regression~\cite{udrescu2020ai} can help to illuminate the underlying physical principles captured by AI models. Unlike standard regression, which fits data to a predefined equation (e.g., a line or polynomial), symbolic regression explores a vast space of mathematical expressions to discover the optimal functional form that describes the data.
Instead of a black box model, we obtain a simple, human-understandable analytical equation that can reveal novel physical correlations or even approximate underlying physical laws.
}

\section{Conclusions}

In this review, we have highlighted selected recent advances in AI-driven nanoscale heat conduction and radiation. We began with machine learning predictions of phonon properties, including phonon dispersion and scattering. We then explored machine learning interatomic potentials and their application to thermal transport in both bulk and interfacial materials. Next, we discussed AI approaches to radiative heat transfer, including solving Maxwell’s equations and the radiative transfer equation, as well as accelerating the inverse design of thermal radiative devices. Finally, we presented open challenges and promising future directions—focusing on data, generalization, uncertainty quantification, and interpretability—that we see as robust opportunities to continue the embrace of AI into thermal transport research.

\bibliography{main.bib}

\end{document}